\documentclass[journal]{IEEEtran} 				
\usepackage{ushort}
\usepackage[mathcal]{euscript}
\usepackage{amsmath}
\usepackage{amsthm}
\usepackage{amssymb}
\usepackage{graphicx}
\usepackage{epsfig}
\usepackage{booktabs}
\usepackage{colortbl}
\usepackage{amsmath}
\usepackage{xcolor}
\usepackage{graphicx}
\usepackage{algorithmic}
\usepackage{algorithm}
\usepackage{cite}
\usepackage{subfigure}
\usepackage{url}

\providecommand{\mbf}[1]{\mathbf{#1}}						
\providecommand{\wt}[1]{\widetilde{#1}}					
\providecommand{\mbft}[1]{\tilde{\mathbf{#1}}}			
\providecommand{\mbfwt}[1]{\wt{\mathbf{#1}}}				

\providecommand{\mc}[1]{\mathcal{#1}}						

\colorlet{tableheadcolor}{gray!25} 
\newcommand{\headcol}{\rowcolor{tableheadcolor}} %
\colorlet{tablerowcolor}{gray!10} 
\newcommand{\rowcol}{\rowcolor{tablerowcolor}} %
\newcommand{\topline}{\arrayrulecolor{black}\specialrule{0.1em}{\abovetopsep}{0pt}%
            \arrayrulecolor{tableheadcolor}\specialrule{\belowrulesep}{0pt}{0pt}%
            \arrayrulecolor{black}}
\newcommand{\midline}{\arrayrulecolor{tableheadcolor}\specialrule{\aboverulesep}{0pt}{0pt}%
            \arrayrulecolor{black}\specialrule{\lightrulewidth}{0pt}{0pt}%
            \arrayrulecolor{white}\specialrule{\belowrulesep}{0pt}{0pt}%
            \arrayrulecolor{black}}

\IEEEoverridecommandlockouts

\begin{document}

\title{Radar Precoder Design for Spectral Coexistence with Coordinated Multi-point (CoMP) System}


\author{Jasmin~A.~Mahal,~Awais~Khawar,~Ahmed~Abdelhadi,~and~T.~Charles~Clancy

\thanks{Jasmin A. Mahal (jmahal@vt.edu), Awais Khawar (awais@vt.edu), Ahmed Abdelhadi (aabdelhadi@vt.edu), and T. Charles Clancy (tcc@vt.edu) are with Virginia Polytechnic Institute and State University, Arlington, VA, 22203.

This work was supported by Defense Advanced Research Projects Agency (DARPA) under the SSPARC program. Contract Award Number: HR0011-14-C-0027. The views, opinions, and/or findings contained in this article/presentation are those of the author(s)/presenter(s) and should not be interpreted as representing the official views or policies of the Department of Defense or the U.S. Government.

Approved for Public Release, Distribution Unlimited.
}

}

\maketitle

\begin{abstract}
This paper details the design of precoders for a MIMO radar spectrally coexistent with a MIMO cellular network. We focus on a coordinated multi-point (CoMP) system where a cluster of base stations (BSs) coordinate their transmissions to the intended user. The radar operates in two modes, interference-mitigation mode when it avoids interference with the CoMP system and cooperation mode when it exchanges information with it. Using either the conventional Switched Null Space Projection (SNSP) or the newly proposed Switched Small Singular Value Space Projection (SSSVSP), the radar beam sweeps across the BS clusters focusing on the optimal ones, optimal in either nullity or difference between the precoded and original radar signal. Taking the channel estimation error into account, the design of precoder is pivoted on the minimal radar interference at the BS clusters during interference-mitigation mode and minimal bit-error-rate at the BSs during cooperation mode while interfering minimally with the radar target 
detection capability. Our investigation shows that loss in radar performance can be compensated using SSSVSP instead of SNSP to some extent but increasing the number of radar antenna elements goes a long way to serve the purpose. Simulations verify our theoretical predictions about the proposed SSSVSP.   
\end{abstract}


\section{Introduction}
Cellular network operators are predicting a $1000\times$ increase in capacity to keep pace with the tremendous growth of aggregate and per-subscriber data traffic \cite{url_1}. Increased investments in infrastructure, e.g. larger number of small cells, and more spectrally efficient technologies, e.g. LTE-Advanced, can help meet this challenge partially.
The Federal Communications Commission (FCC) is considering a number of options including incentive auctioning and sharing of federal spectrum to meet the commercial spectrum requirements. Of these two, spectrum sharing is a quite promising technology due to the large number of under-utilized federal spectrum bands that can be shared with commercial cellular operators to satisfy their growing demands. But spectrum sharing is associated with its inherent set of challenges because the incumbents need to be protected from the harmful interference that can arise due to the operation of other systems in the shared bands. In this paper, the authors have addressed the specific issue of spectrum sharing between a multiple-input multiple-output (MIMO) radar system and a commercial MIMO cellular network consisting of clusters of cooperative base stations (BS), commonly known as coordinated multi-point system.\\ 
 
\indent \textbf{Coordinated Multi-point (CoMP) Systems:} MIMO is a key enabling technique for improving the throughput of future wireless broadband system. Despite its theoretical attractiveness, a MIMO commercial cellular system is essentially interference-limited because the transmission in each cell causes interference to the other cells, i.e. inter-cell interference (ICI). For cell-edge user equipments (UEs), ICI is severe due to their proximity to a number of neighboring cells. Downlink is more interference-limited than the uplink because the sophisticated interference suppression techniques can be easily implemented in BSs but they are not practical for UEs. Implementation of these techniques would render the UEs power-inefficient or incompact. Moreover, coordination among UEs to mitigate interference is challenging and, thus, does not alleviate the ICI problem. Conventional approaches for mitigating the ICI, such as static frequency reuse, spread spectrum, and sectoring etc. are not efficient for MIMO 
networks owing to their respective inherent limitations \cite{Andrews2007}. Due to the recent advancements in processing capabilities at BSs and the increase in backhaul capacity, coordination among BSs can be utilized to mitigate the ICI. Such a configuration is known as coordinated multi-point (CoMP) transmission/reception which not only controls the ICI but also takes advantage of it. Due to the advantages of the CoMP system over traditional cellular system it has been included in the fourth generation (4G) mobile standard, i.e., 3GPP LTE-Advanced Release 11 and beyond consider CoMP as an enabling technology for 4G mobile systems \cite{LTE_Rel11}. In short, the CoMP system coordinates simultaneous transmissions from multiple BSs to UEs in the downlink and perform joint decoding of UE signals at multiple BSs in the uplink. This results in improved coverage, throughput, and efficiency for cellular system.\\ 

\indent \textbf{Regulatory Efforts to Share Radar Spectrum:} In 2010, the U.S. President Obama issued a Presidential Memorandum which recommended the Federal Government to make available 500 MHz of Federal or non-federal spectrum for both mobile and fixed wireless broadband use by commercial users within 10 years \cite{Memo2010}. In compliance with the Presidential directive, the National Telecommunications and Information Administration (NTIA) identified a number of bands compatible for commercial utilization including the 3500-3650 MHz band which is primarily used by federal radar systems \cite{NTIA10}. However, the NTIA also pointed out in a study that if the identified radar bands are shared with commercial wireless services, exclusion zones are required to protect wireless services and these exclusion zones may go upto hundreds of kilometers \cite{NTIA12}. On the other hand, the Federal Communications Commission (FCC) has proposed to use small cells that will be sharing the 3.5 GHz band with the federal 
radar systems \cite{FCC12_SmallCells}. In the past, government bands have been successfully shared with commercial wireless systems. Example includes: Wi-Fi and Bluetooth in the 2450-2490 MHz band and wireless local area network (WLAN) in the 5.25-5.35 and 5.47-5.725 GHz radar bands \cite{FCC_5GHz_Radar06}.

\subsection{Related Work}
Various schemes have been proposed to share radar spectrum with communication systems. Spectrum sharing schemes based on waveform shaping was first proposed by Sodagari et. al. \cite{S.SodagariDec.2012} in which the radar waveform was shaped in a way so that it was in the null space of interference channel between the radar and a communication system. This scheme was extended to multi-cell communication systems by Khawar et. al. \cite{A.Khawar} in which spectrum was shared between a MIMO radar and a LTE cellular system with multiple base stations. They proposed algorithms to select the interference channel with the maximum null space to project the radar signal onto that null space. Although this paper addressed a more practical scenario with multiple BSs, at any given time it mitigated interference to only one BS, i.e, the station with the maximum nullity. A similar approach was presented by Babaei et. al. in which radar waveform was shaped to mitigate interference to all the BSs in the network \cite{
Babaei2013}. Other spectrum sharing approaches include cooperative sensing based spectrum sharing where a radar's allocated bandwidth is shared with communication systems \cite{WMW+08, BNR12, SPC12, GhorbanzadehMilcom2014, SKA+14DySPAN}, joint communication-radar platforms that are spectrally-agile \cite{FHH14}, database-aided sensing at communication systems to enable radar spectrum sharing \cite{PMM+14}, radar systems that form virtual arrays to coexist with communications system \cite{Shahriar_WCNC}, and beamforming approaches at MIMO radars \cite{DH13}.


\subsection{Our Contributions}
Building upon the works in \cite{A.Khawar} and \cite{Babaei2013}, the authors of this paper have extended the solution approach to address a MIMO radar and a commercial CoMP communication system coexistence scenario, which is applicable for the LTE-Advanced system. Our contributions are summarized as follows:

  
\begin{itemize}
\item \textbf{Precoder Design for Interference Mitigation:} In order to mitigate interference to CoMP system we present two novel radar precoder design methods in this paper. In the first scheme, the radar projects its signals onto the clusters formed by the CoMP system and selects the cluster with the maximum null space. This scheme is known as Switched Null Space Projection (SNSP) as radar looks for the optimal cluster at each pulse repetition interval (PRI) and thus switches among clusters at each PRI depending upon its mission requirements. In the second scheme, called Switched Small Singular Value Space Projection (SSSVSP), the projection space has been expanded to include the subspace corresponding to the small non-zero singular values under a specified threshold in addition to the null space. The precoder is designed based on the knowledge of a composite interference channel matrix which comprises of the channel matrices from radar to all the BSs in a particular cluster. Based on the proposed channel 
estimation technique in \cite{A.BabaeiJuly2013}, all of the BSs in the cluster coordinate in their choices of training symbols and power transmission during the channel estimation phase and the radar transmitter can estimate the composite interference channel between them.

\item \textbf{Precoder Design for Cooperation:} We design a novel radar precoder for communicating with the CoMP system. The purpose of this precoder is information exchange between radar and the CoMP system. The CoMP system primarily informs the radar of the clustering information and in return radar informs the CoMP system about which cluster it has selected for interference mitigation. Basically, this precoder is vital for the success of interference mitigation techniques. Without this sort of information exchange, spectrum sharing would not be successful. We design simple linear precoders based on zero forcing (ZF) and minimum mean-square-error (MMSE) criteria.




\item \textbf{Precoder Performance Analysis:} In order to evaluate the performance of radar precoders, we perform simulations in the presence of channel estimation errors \cite{KAC14_Milcom}. We look at the target localization and interference mitigation capabilities of the radar precoders. The results indicate that while the precoder nullifies the radar interference to the clusters, it degrades the radar performance by introducing correlation in the probing signals. We show that this performance loss can be compensated for by two means either by increasing the number of radar antennas or by utilizing small singular value space projection rather than using null space projection. Our results show that between the two compensation schemes, the former is more effective. 

\end{itemize}

 

\subsection{Notations}
The vectors and matrices are denoted by lower-case and uppercase boldface letters, respectively (e.g., $\textbf{j}$ and $\textbf{J}$). The rank, null space, transpose, and Hermitian transpose of $\textbf{J}$ are represented by rank$\left\{\textbf{J}\right\}$, $\mathcal{N}\left\{\textbf{J}\right\}$, ${\textbf{J}}^T$ and ${\textbf{J}}^*$ , respectively. The subspace spanned by a set of vectors $S$ is denoted by $\text{Span}\left\{S\right\}$. The $N$ by $N$ identity matrix is presented by ${\textbf{I}}_N$. $\left\|\cdot\right\|_2$ and $\left\|\cdot\right\|_F$ denote L-2 norm and Frobenius norm respectively. For a quick reference, important notations are summarized in Table \ref{tab:Notations}.


\renewcommand{\arraystretch}{1.5}
\begin{table}
\centering
\caption{Table of Notations}
\begin{tabular}{ll}
  \topline
  \headcol Notation &Description\\
  \midline
			$M$						&Total number of base stations (BSs) \\
\rowcol			$K$						&Total number of user equipments (UEs) \\
			$\mc M_k$				&Cluster of BSs serving $k^{\text{th}}$ UE \\
\rowcol			$M_k$				&Total number of BSs in cluster $\mc M_k$ \\

			$M_R$ 						&Radar transmit/receive antennas \\
\rowcol     $N_{\text{BS}}$				&BS transmit/receive antennas  \\ 
			$N_{\text{UE}}$				&UE transmit/receive antennas  \\ 

	\rowcol	$\mbft H_{R,i}^k$			&Composite interference channel between radar\\ 
	                              &and the $i^{\text{th}}$ cluster $\mc M_k$\\
%
	
  \hline
\end{tabular}
\label{tab:Notations}
\end{table}

\subsection{Organization}
The remainder of the paper is structured as follows. Section \ref{sec:models} details the radar/communication system spectral-coexistence model. Design of radar precoder for different operating modes is described in Section \ref{sec:precod}. The newly proposed SSSVSP algorithms are presented in Section \ref{sec:algos}. Section \ref{sec:sim} explains the simulation results. Section \ref{sec:conc} concludes the paper with final remarks.

\section{Radar/CoMP System Spectral-Coexistence Models}\label{sec:models}

In this section, we introduce CoMP communication system model, MIMO radar signal model, and our radar-communication system spectrum sharing scenario. 

\subsection{Coordinated Multipoint (CoMP) System}
We consider a CoMP system with a total of $M$ base stations forming a set of BSs, denoted by $\mathcal{M}$, and a total of $K$ users forming a set of UEs, denoted by $\mathcal{K}$. Each BS is equipped with $N_{\text{BS}}$ antennas and each UE is equipped with $N_{\text{UE}}$ antennas. 
The $k^\mathrm{th}$ UE receives its message from a cluster of $M_k$ base stations $\mathcal{M}_k \subseteq \mathcal{M}$ or the $m^\mathrm{th}$ BS is provided with the messages of its assigned users set $\mc K_m \subseteq \mc K$.
Let ${\mbf{u}_k = [u_{k,1} ...u_{k,D_k} ]^T} \in \mathbb{C}^{D_k}$ be the vector of transmit symbols for the $k^\mathrm{th}$ UE. This complex vector represents $D_k$ independent information streams intended for user $k$, where $D_k \leq min({ M_k} {N_{\text{BS}}}, N_{\text{UE}} )$. $D_k$ is known as the degrees of freedom (DoF) and is defined as the number of signalling dimensions, where one signaling dimension is associated with one interference-free information stream \cite{Negro2010}. All the BSs in the cluster $\mathcal{M}_k$ are informed about the data streams $\mbf{u}_k$.  This can be realized by utilizing the backhaul links among the BSs and the central switching unit. Assuming $m \in \mathcal{M}_k $, the $m^\mathrm{th}$ BS maps the vector $\mbf{u}_k$ via a matrix $\mbf{F}_{k,m} \in \mathbb{C}^{ N_{\text{BS}} \times D_k} $ onto the signal $\tilde{\mbf{z}}_m \in \mathbb{C}^ {N_{\text{BS}}}$ transmitted by the $m^\mathrm{th}$ BS  which can be expressed as \cite{S.KavianiDec.2011}
\begin{equation}
\tilde{\mbf{z}}_m=\sum_{k\in \mathcal{K}_m} \mbf{F}_{k,m} \mbf{u}_k. 
\end{equation}
Assuming $P_m$ to be the power constraint of the $m^\mathrm{th}$ BS, the following constraint has to be satisfied,
\begin{align}
E\left[\left\|{\tilde{\textbf{z}}_m}^2\right\|\right]&=tr\left\{E\left[\tilde{\textbf{z}}_m {\tilde{\textbf{z}}^{*}_{m}}\right]\right\}\\ \notag
&=\sum_{k\in \mathcal{K}_m} tr\left\{\textbf{F}_{k,m} \textbf{F}^{*}_{k,m}\right\} \leq P_m, m=1,\cdots,M_k.
\end{align}

The signal received by the $k^\mathrm{th}$ user on the downlink is given by 
\begin{align}
\label{eq:net-mimo_1}{\mbf{y}}_k
&=\sum_{m\in \mathcal{M}_k} \tilde{\mbf{G}}_{k,m} {{\mbf{F}}_{k,m}} {{\mbf{u}}_{k}}
+{\tilde{\mbf{n}}_{k}}
\end{align}
where $\tilde{\mbf{G}}_{k,m} \in \mathbb{C}^{ N_{\text{UE}} \times N_{\text{BS}}}$ is the channel matrix between the $m^\mathrm{th}$ BS and $k^\mathrm{th}$ user and $\tilde{\mbf{n}}_{k}$ accounts for noise and non-radar interference terms. 

Assuming joint detection mode among the BSs on the uplink and further assuming reciprocity of the channel, the signal received by the $m^\mathrm{th}$ BS in a cluster on the uplink is given by 
\begin{equation}
\label{eq:net-mimo_2}{\mbf{b}}_m=\sum_{k \in \mc K_m} {\tilde{\mbf{G}}}^*_{k,m} {\tilde{\mbf{a}}_{k}}+{\tilde{\mbf{n}}_{m}} 
\end{equation}  
where $\tilde{\mbf{a}}_{k} \in \mathbb{C}^{ N_{\text{UE}} \times 1}$ is the signal vector transmitted by the $k^\mathrm{th}$ UE and $\tilde{\mbf{n}}_{m}$ accounts for noise and non-radar interference terms.

\subsection{Clustering Algorithms}

Joint transmission/reception in the CoMP system requires additional signalling overhead and robust backhaul channels. Due to this reason, only a limited number of BSs cooperate to form a cluster \cite{Boccardi2007}. Cluster formation is an important aspect of the CoMP system in order to exploit benefits promised by the CoMP system. In general, static and dynamic clustering algorithms have been proposed that may or may not have overlapping. Each scheme has its own merits which are discussed as follows \cite{MF11}:
\begin{itemize}
	\item \textbf{Static Clustering:} Static clusters do not change over time and are designed based on the time-invariant network parameters such as the geography of BSs and surroundings. In static BS clustering algorithms, the pre-specified BS clusters do not change with time and as the neighboring BSs interfere the most with each other on average, the clusters are formed by the adjacent BSs only \cite{Boccardi2007}. This leads to very limited performance gains as the changing channel conditions are not taken into account.
	\item \textbf{Dynamic Clustering:} Dynamic clusters continuously adapt to changing parameters in the network such as UE locations and RF conditions. Dynamic clustering schemes exploit the effect of changing channel conditions, which leads to much higher performance gains without a significant overhead increase \cite{Papadogiannis2008}, \cite{Boccardi2008}. So the dynamic clustering algorithms cluster the BSs not based on their geographical proximity  but rather based on their level of interference in the absence of any cooperation \cite{Papadogiannis2010}. Thus, by capturing the effect of changing channel conditions, dynamic clustering algorithms exploit the inherent macrodiversity of multicell wireless systems \cite{Papadogiannis2008}.
	\item \textbf{Non-Overlapping vs Overlapping Clusters:} Static or dynamic clustering algorithms can be based on schemes which do not support overlap, i.e. clusters are disjunct with respect to the cells involved. However, such a scheme may result in a low signal-to-interference-noise-ratio (SINR) for a UE at the border between clusters. This issue can be addressed by having spatially overlapping clusters. Although, overlapping clusters result in better performance in terms of SINR at a UE, the price paid is the loss of multi-user diversity in each cluster \cite{MF11}.

\end{itemize}
In order to keep precoder design simple and tractable we limit ourselves to the case where the CoMP system use static non-overlapping clustering algorithms where the clusters are formed by adjacent cells which is useful for radar since it will illuminate a specific geographical area. 
  We have assumed that mobile network operators are performing the task of BS clustering and the clustering information is communicated to the radar during the cooperation mode.

\subsection{Colocated MIMO Radar}
MIMO radar is an emerging area of research and a possible upgrade option of legacy radar systems. Unlike the standard phased-array radar which transmits scaled versions of a single waveform, MIMO radars transmit multiple probing signals that can be chosen freely. This diversity in waveform enables superior capabilities when compared with a standard phased-array radar. In this paper, we have considered a collocated MIMO radar with $M_R$ antenna elements. If we denote the samples of baseband equivalent of $M_R$-dimensional transmitted radar signals as $\left\{\mbf{x}_R(n)\right\}^{L}_{n=1}$, the signal coherence matrix 
can be written as \cite{KAC+14ICNC}
 
\begin{equation}
\mbf{R}_{\mbf x}=\frac{1}{L} \sum^L_{n=1} \mbf{x}_R\left(n\right) {{\mbf{x}}^{*}_R}\left(n\right) =
\begin{bmatrix}
1 & \beta_{12} & \ldots & \beta_{1{M_R}} \\
\beta_{21} & 1 &  \ldots & \beta_{2{M_R}} \\
\vdots & \vdots & \vdots & \vdots \\
\beta_{{M_R}1} & \beta_{{M_R}2} &  \ldots & 1 \\
\end{bmatrix}
\end{equation}
where $n$ is the time index, $L$ is the total number of time samples and $\beta_{oc}$ denotes the correlation coefficient between $o^\mathrm{th}$ and $c^\mathrm{th}$ signals $\left(1 \leq o,c \leq M_R \right)$. The phases of $\left\{\beta_{oc}\right\}$ direct the beam to the angle of interest. If $\beta_{oc} = 0$ for $o \neq c$, then $\mbf{R}_{\mbf x} = \mbf{I}_{M_R}$, i.e, orthogonal waveforms. This corresponds to omni-directional transmission.

\begin{figure}[!t]
\centering
\includegraphics[width=0.45\textwidth]{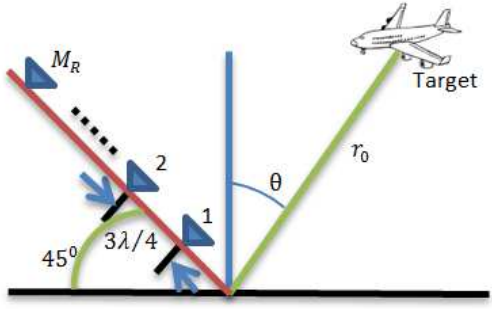}
\caption{Configuration of radar antenna array}
\label{fig_arrconfig}
\end{figure}

The signal received from a single point target at an angle $\theta$ can be written as \cite{KAC14_TDetect}
\begin{equation}
\mbf y(n) = \alpha \, \mbf A(\theta) \,  {\mbf {x}}_R(n) + \mbf w(n) \label{eqn:rxRadar}
\end{equation}
where $\alpha$ represents the complex path loss including the propagation loss and the coefficient of reflection and $\mbf A (\theta)$ is the transmit-receive steering matrix defined as
\begin{equation}
\mbf A (\theta) \triangleq \mbf a_t(\theta) \mbf a_r^T(\theta).
\end{equation} 

\noindent Denoting the propagation delay between the target and the $p^{\mathrm{th}}$ transmit element by $\tau_{t,p}(\theta)$ and the delay between the target and the $l^{\mathrm{th}}$ receive element by $\tau_{t,l}(\theta)$, the total propagation delay between the $p^{\mathrm{th}}$ transmit element and the $l^{\mathrm{th}}$ receive element is given by

\begin{equation}
	\tau_{pl}(\theta)=\tau_{t,p}(\theta) + \tau_{t,l}(\theta).
\end{equation}
Based on these  notations of propagation delays, the transmit steering vector, $\mbf a_t(\theta)$ is defined as
\begin{equation}
\mbf a_t(\theta) \triangleq \begin{bmatrix} e^{-j \omega_c \tau_{t,1}(\theta)} &e^{-j \omega_c \tau_{t,2}(\theta)} &\cdots &e^{-j \omega_c \tau_{t,{M_R}}(\theta)} \end{bmatrix}^T,
\label{eq:at}
\end{equation}
and the receive steering vector, $\mbf a_r(\theta)$ is defined as
\begin{equation}
\mbf a_r(\theta) \triangleq \begin{bmatrix} e^{-j \omega_c \tau_{t,1}(\theta)} &e^{-j \omega_c \tau_{t,2}(\theta)} &\cdots &e^{-j \omega_c \tau_{t,{M_R}}(\theta)} \end{bmatrix}^T.
\label{eq:ar}
\end{equation}
Using this model, the Cram\'er Rao Bound (CRB) for target direction estimation is given in equation \eqref{eq:crb} \cite{LS08}. In equation \eqref{eq:crb}, ${{\dot{\textbf{a}}}_t}\left(\theta\right)=\frac{d{\textbf{a}}_t}{d\theta}$ and ${{\dot{\textbf{a}}}_r}\left(\theta\right)=\frac{d{\textbf{a}}_r}{d\theta}$. Assuming all other parameters fixed, the performance of target direction estimation in terms of maximum likelihood or Cram\'er Rao bound is optimal for orthogonal probing signals \cite{LS08, KAC14DySPANWaveform, KAC14_QPSK}.
\begin{figure*}
\hrulefill
\begin{eqnarray}
\label{eq:crb}
\mathrm{CRB}(\theta) = \frac{1}{2\: \text{SNR}} \left(M_R \dot{\mathbf{a}}_t^*(\theta)\mathbf{R}_{\mbf x}^T\dot{\mathbf{a}}_t(\theta) + \mathbf{a}_t(\theta)^* \mathbf{R}_{\mbf x}^T \mathbf{a}_t(\theta) \|\dot{\mathbf{a}}_r(\theta)\|^2 - \frac{M_R\left|\mathbf{a}_t^*(\theta) \mathbf{R}_{\mbf x}^T \mathbf{\dot{a}}_t(\theta)\right|^2}{\mathbf{a}_t^*(\theta) \mathbf{R}_{\mbf x}^T \mathbf{a}_t(\theta)}\right)^{-1}
 \end{eqnarray}
\hrulefill
\end{figure*}

\subsection{Spectral Coexistence Scenario}
\begin{figure}[!t]
\centering
\includegraphics[width=0.45\textwidth]{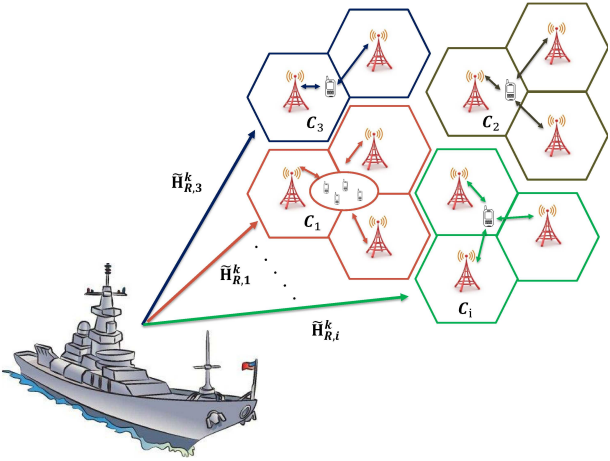}
\caption{Spectral coexistence of a MIMO radar with a coordinated multi-point (CoMP) system. An uplink-downlink model of the CoMP system while sharing radar bands is illustrated.}
\label{fig_cluster}
\end{figure}

The communication system shares the spectrum with a monostatic ship-borne MIMO radar system as shown in Fig. (\ref{fig_cluster}). 
The composite interference channel between radar transmitter and the $i^{\text{th}}$ cluster ${\mathcal{M}}_k$ is denoted by $\tilde{\mbf{H}}^k_{R,i}  \in \mathbb{C}^ {{M_k}{N_{\text{BS}}} \times M_R} $. The superscript $k$ implies that the cluster$\text{'}$s message is intended for UE $k$ and $1 \leq i\leq C_T$ where $C_T$ is the total number of clusters in the CoMP system. 
Channels are assumed to be block faded and quasi-static. The signal received by the $i^{\text{th}}$ BS cluster ${\mathcal{M}}_k$ on the uplink, in the presence of radar, is given simply by 
\begin{equation} \label{eqH}
{\mbf{b}}_{C_i}=\sum_{m \in C_i} \sum_{k \in \mc K_m} {\tilde{\mbf{G}}}^*_{k,m} {\tilde{\mbf{a}}_{k}}+ \tilde{\mbf{H}}^k_{R,i} \mbf x_R + {\tilde{\mbf{n}}_{m}} 
\end{equation}  
where $\tilde{\mbf{H}}^k_{R,i} \mbf x_R$ is the interfering signal from the MIMO radar which we want to mitigate by designing precoders in the next section.

%


\section{Radar Precoder Design}\label{sec:precod}
In this paper, according to (\ref{eqH}), only the $\mbf{H}$ matrix would play a role in the design of radar precoder but not the $\mbf{G}$ or $\mbf{F}$ matrices. The goal is to construct the radar precoded signal $\mbfwt{x}_R$ such that its interference at all of the BSs of the optimally chosen cluster of the CoMP system is either zero-forced or minimized during interference-free mode and the radar keeps switching the beam across the optimal clusters with time. On the other hand, for cooperation mode when radar is communicating information to the BSs, we have to make sure that the bit-error-rate (BER) of the received signal at the BSs is low enough for effective detection.

\subsection{Precoder Design for Interference Mitigation}
Radar interference to the $i^\mathrm{th}$ cluster $\mathcal{M}_k$ can be mitigated by designing the radar precoding matrix ${\mbf{P}_{R,i}}$ as\begin{eqnarray}
\mbf H^k_{nR,i} \mbf P_{R,i} \mbf x_R &=& 0, \quad \forall n\in \mathcal{M}_k,\\
\mbf H^k_{nR,i} \mbfwt x_R &=& 0,  \quad  \forall n\in \mathcal{M}_k.
\end{eqnarray}
The above criteria is satisfied by the following condition
\begin{equation}\label{eqn:19}
\mbf P_{R,i} \mbf x_R \in \mathcal{N}\left(\mbf H^k_{nR,i}\right), \quad \forall n\in \mathcal{M}_k.
\end{equation}
Therefore, the transmitted radar signal must lie in the null space of ${\mbf{H}^k_{nR,i}}$ for all $n$, where $n=1,2,\cdots,M_k$. We can rewrite equation \eqref{eqn:19} as
\begin{equation}
{\mbf{P}_{R,i}} {\mbf{x}}_R \in \mathcal{N}\left({\mbf{H}^k_{1R,i}}\right) \cap \mathcal{N}\left({\mbf{H}^k_{2R,i}}\right)\cdots \cap \mathcal{N}\left({\mbf{H}^k_{M_k R,i}}\right).
\end{equation}
Using the equality $\mathcal{N} \left(\mbf{A}\right) \cap \mathcal{N} \left(\mbf{B}\right)=\mathcal{N} \left(\mbf{C}\right)$ where $\mbf{C}={\left[\mbf{A}^*  \mbf{B}^*\right]}^*$, the precoder must satisfy that
\begin{equation}
{\mbf{P}_{R,i}} {\mbf{x}}_R \in \mathcal{N}\left(\tilde{\mbf{H}}^k_{R,i}\right)
\end{equation}
where
\begin{equation}
\tilde{\mbf{H}}^k_{R,i}={\left[ {\left({\mbf{H}^k_{1R,i}}\right)}^*  {\left({\mbf{H}^k_{2R,i}}\right)}^*\cdots{\left({\mbf{H}^k_{M_k R,i}}\right)}^* \right]}^*.
 \end{equation} 
To find the null space, we have to first obtain the Singular Value Decomposition (SVD) of direct communication link \cite{CadambeAug2008} which diagonalizes the channel matrix ${\tilde{\mbf{H}}}^k_{R,i} \in \mathbb{C}^{M_k N_{\text{BS}}\times M_R}$. The regular form of the SVD of ${\tilde{\mbf{H}}}^k_{R,i}$ is as follows
\begin{equation}
	\tilde{\mbf{H}}^k_{R,i}=\tilde{\mbf{U}}_i \tilde{\mbf{S}}_i {\tilde{\mbf{V}}^*_i}
\end{equation}
where $\tilde{\mbf{U}}_i \in \mathbb{C}^ {{M_k}{N_{\text{BS}}} \times {M_k}{N_{\text{BS}}}}$ is a unitary matrix, $\tilde{\mbf{S}}_i \in \mathbb{C}^ {{M_k}{N_{\text{BS}}} \times {M_R}}$ is a rectangular diagonal matrix with non-negative real numbers on the diagonal, and $\tilde{\mbf{V}}^*_i \in \mathbb{C}^ {{M_R} \times {M_R}}$ is a unitary matrix. The diagonal entries of $\tilde{\mbf{S}}_i$ are known as the singular values of $\tilde{\mbf{H}}^k_{R,i}$. The ${M_k}{N_{\text{BS}}}$ columns of $\tilde{\mbf{U}}_i$ and the ${M_R}$ columns of $\tilde{\mbf{V}}_i$ are called the left-singular vectors ($\mbf{u} \in \mathbb{C}^ {{M_k}{N_{\text{BS}}} \times 1}$) and right-singular vectors ($\mbf{v} \in \mathbb{C}^ {{M_R} \times 1}$) of $\tilde{\mbf{H}}^k_{R,i}$, respectively.

\begin{figure*}
\hrulefill
\begin{align}\label{eq:A1}
\tilde{\mbf{H}}^k_{R,i}
&=\underbrace{\begin{bmatrix} \mathbf{u}_1 & \cdots & \mathbf{u}_q &|& \mathbf{u}_{q+1} & \cdots & \mathbf{u}_{M_k N_{\text{BS}}} \end{bmatrix}}_{\tilde{\mbf{U}}_i} \underbrace{{\left[\begin{array}{c|c} \mathbf{\Sigma}_{q,q} & \mbf{0}_{q,(M_R-q)} \\ \hline \mbf{0}_{(M_k N_{\text{BS}}-q),q} & \mbf{0}_{(M_k N_{\text{BS}}-q),(M_R-q)} \end{array}\right]}}_{\tilde{\mbf{S}}_i} \underbrace{{\left[\begin{array}{c} \mathbf{v}^*_1\\ \vdots \\ \mathbf{v}^*_q \\ \hline  \mathbf{v}^*_{q+1} \\ \vdots \\ \mathbf{v}^*_ {M_R}\end{array}\right]}}_{\tilde{\mbf{V}}^*_i}\\ \notag \label{eq:A2}
&=\begin{bmatrix} \mathbf{u}_1 & \cdots & \mathbf{u}_q \end{bmatrix} \underbrace{\begin{bmatrix} \sigma_1 & {\cdots} & {0}\\ {\vdots}&\ddots & {\vdots}\\{0}&{\cdots}&\sigma_q \end{bmatrix}}_{\mbf{\Sigma}_{q,q}} {\left[\begin{array}{c} \mathbf{v}^*_1\\ \vdots \\ \mathbf{v}^*_q \end{array}\right]}+
\begin{bmatrix} \mathbf{u}_{q+1} & \cdots & \mathbf{u}_{M_k N_{\text{BS}}} \end{bmatrix} \left[\begin{array}{c} \mbf{0}_{(M_k N_{\text{BS}}-q),(M_R-q)} \end{array}\right] \underbrace{{\left[\begin{array}{c} \mathbf{v}^*_{q+1} \\ \vdots \\ \mathbf{v}^*_{M_R} \end{array}\right]}}_{\bar{\mbf{V}}^*_i} \\&\quad \quad \quad +\begin{bmatrix} \mathbf{u}_1 & \cdots & \mathbf{u}_q \end{bmatrix} \left[\begin{array}{c} \mbf{0}_{q,(M_R-q)} \end{array}\right] \left[\begin{array}{c} \mathbf{v}^*_{q+1} \\ \vdots \\ \mathbf{v}^*_{M_R} \end{array}\right] + \begin{bmatrix} \mathbf{u}_{q+1} & \cdots & \mathbf{u}_{M_k N_{\text{BS}}} \end{bmatrix} \left[\begin{array}{c} \mbf{0}_{(M_k N_{\text{BS}}-q),q} \end{array}\right]{\left[\begin{array}{c} \mathbf{v}^*_1\\ \vdots \\ \mathbf{v}^*_q \end{array}\right]}\\
&=\begin{bmatrix} \mathbf{u}_1 & \cdots & \mathbf{u}_q \end{bmatrix} {\begin{bmatrix} \sigma_1 & {\cdots} & {0}\\ {\vdots}&\ddots & {\vdots}\\{0}&{\cdots}&\sigma_q \end{bmatrix}} {\left[\begin{array}{c} \mathbf{v}^*_1\\ \vdots \\ \mathbf{v}^*_q \end{array}\right]}\label{eq:A3} 
\end{align}
\hrulefill
\end{figure*}
 The partitioned matrices and the outer product form of the SVD of ${\tilde{\mbf{H}}}^k_{R,i}$ is as shown in equations (\ref{eq:A1}) through (\ref{eq:A3}) \cite{Kalm1996}. These partitions assume that the total number of non-zero singular values, $q$ is strictly less than $M_k N_{\text{BS}}$ and $M_R$. From equation (\ref{eq:A2}), it is obvious that only the first term, i.e. the first $q$ left singular vectors, $\mbf{u}_1 \cdots \mbf{u}_q$, and right singular vectors, $\mbf{v}_1 \cdots \mbf{v}_q$, make any contributions to $\tilde{\mbf{H}}^k_{R,i}$. The rest of the three terms are zero. Among these three terms, the first term is associated with the null space corresponding to the zero singular values. So, for null space projection, the radar beam is projected onto the space formed by the second term of equation (\ref{eq:A2}) which has no contributions to $\tilde{\mbf{H}}^k_{R,i}$. Using SVD, null space of $\tilde{\mbf{H}}^k_{R,i}$ is $\text{Span}\left\{\bar{\mbf{V}}_i\right\}$ where $\bar{\mbf{V}}_i$ 
comprises of the columns of ${\tilde{\mbf{V}}_i}$ corresponding to zero singular values of $\tilde{\mbf{H}}^k_{R,i}$. As a result, the precoder $\mbf{P}_{R,i}$ must be the projection matrix into $\text{Span}\left\{\bar{\mbf{V}}_i\right\}$, 
\begin{equation}
\mbf{P}_{R,i}=\bar{\mbf{V}}_i {\bar{\mbf{V}}}_i^*.
\end{equation}

Although this precoder will eliminate the radar interference to all the BSs in the $i^{\text{th}}$ cluster $\mathcal{M}_k$, it will affect the spatial correlation of radar probing signals and change their coherence matrix. For precoded radar signals, the coherence matrix, $\mbf{R}_{\mbf x,i}=\mbf P_{R,i} \mbf P_{R,i}^*$. 

Assuming that the elements of the channel matrices are independent and identically distributed (i.i.d.) and drawn from a continuous distribution, by random matrix theory, the rows of $\tilde{\mbf{H}}^k_{R,i}$ are linearly independent \cite{Jafar2011}. The matrix $\tilde{\mbf{H}}^k_{R,i}$ is therefore full rank. So, the nullity, i.e. the dimension of null space of $\tilde{\mbf{H}}^k_{R,i}$ is greater than the difference metric, $\left(M_R - M_k N_{\text{BS}}\right)$. In mathematical terms,
\begin{equation}
	\text{null} \,\left[\tilde{\mbf{H}}^k_{R,i}\right]={\text{dim}}\left[\mathcal{N} \left(\tilde{\mbf{H}}^k_{R,i}\right)\right]=\left(M_R - M_k N_{\text{BS}}\right)^+.
\end{equation}

The necessary condition for a non-trivial precoder $\left(\mbf{P}_{R,i} \neq 0\right)$ to exist is that the number of radar transmit antennas is greater than sum of the requested DoF of all the base stations in a cluster. In that case, we must have,
\begin{equation*}
M_R > M_k N_{\text{BS}}
\end{equation*}
to have a non-zero nullity for $\tilde{\mbf{H}}^k_{R,i}$ and hence a non-zero precoder. 

In SVD, the singular values are always arranged in decreasing order:
\begin{equation}
\sigma_q < \sigma_{q-1} <\cdots < \sigma_1.	
\end{equation}
These singular values are in fact measures of the transmitted power in the directions of their corresponding right singular vectors. Thus directing the radar transmitted power along vectors associated with the small singular values is equivalent to reducing radar interference in those directions. Consequently, projecting the radar transmitted signal onto a space spanned by the right singular vectors corresponding to singular values smaller than a specific threshold, $\sigma_{\mathrm{th}}$, will minimize radar interference to the cluster rather than eliminating them completely. This compromise is expected to mitigate the performance loss in radar target detection capability to some extent. The value of $\sigma_{\mathrm{th}}$ is dependent on the power constraints of the communication system. With, $\sigma<\sigma_{th}$, i.e. when the projection space is formed by the right singular vectors corresponding to the non-zero small singular values under a threshold in addition to the zero singular values, the 
interference would be reduced to a great extent for the current channel condition. This kind of projection could be called Small Singular Value Space Projection (SSVSP). For SSVSP, using SVD of $\tilde{\mbf{H}}^k_{R,i}$, the projection space is $\text{Span}\left\{\bar{\mbf{V}}_{s,i}\right\}$ where $\bar{\mbf{V}}_{s,i}$ comprises of the columns of ${\tilde{\mbf{V}}_i}$ corresponding to the small non-zero singular values under $\sigma_{th}$ in addition to the zero singular values. As a result, the precoder $\mbf{P}_{R_{s},i}$ is given by 

\begin{equation}
\mbf{P}_{R_s,i}=\bar{\mbf{V}}_{s,i}{\bar{\mbf{V}}}^*_{s,i}.
\end{equation} 

Under switched beam condition, the radar keeps scanning for the best composite interference channel between itself and the clusters, and projects its beam towards that cluster. Assuming $C_T$ represents the total number of clusters at a specific time in the CoMP system, with Switched Null Space Projection (SNSP), the best and the worst composite interference channels are selected in the following way \cite{A.Khawar}
\begin{equation}
\tilde{\mbf{H}}_{best}=(\tilde{\mbf{H}}^k_{R,i})_{i_{max}}
\end{equation}
where
\begin{equation}
i_{max}=\underset{1 \leq i\leq C_T}{\text{arg max }} {\text{dim}}\left[\mathcal{N} \left(\tilde{\mbf{H}}^k_{R,i}\right)\right]
\end{equation}
and
\begin{equation}
\tilde{\mbf{H}}_{worst}=(\tilde{\mbf{H}}^k_{R,i})_{i_{min}}
\end{equation}
where
\begin{equation}
i_{min}=\underset{1 \leq i\leq C_T}{\text{arg min }} {\text{dim}}\left[\mathcal{N} \left(\tilde{\mbf{H}}^k_{R,i}\right)\right]
\end{equation}

%
%

\noindent Assuming full rank $\tilde{\mbf{H}}^k_{R,i}$, the nullity of $\tilde{\mbf{H}}^k_{R,i}$ is defined as
\begin{equation}
{\text{dim}}\left[\mathcal{N} \left(\tilde{\mbf{H}}^k_{R,i}\right)\right]=\left(M_R - M_k N_{\text{BS}}\right)^+
\end{equation}
where $M_k$ is the number of BSs in the $i^\mathrm{th}$ cluster $\mc M_k$.
    
For newly proposed Switched Small Singular Value Space Projection (SSSVSP), the best and the worst composite interference channels are chosen as follows:
\begin{equation}\label{eq:crit1}
\tilde{\mbf{H}}_{best}=(\tilde{\mbf{H}}^k_{R,i})_{i_{min}}
\end{equation}
where
\begin{equation}\label{eq:crit2}
 i_{min}=\underset{1 \leq i\leq C_T}{\text{arg min }} \left\|\mbf{P}_{{R_s},i} \mbf{x}_R-\mbf{x}_R\right\|_2
\end{equation}
and
\begin{equation}
\tilde{\mbf{H}}_{worst}=(\tilde{\mbf{H}}^k_{R,i})_{i_{max}}
\end{equation}
where
\begin{equation}
i_{max}=\underset{1 \leq i\leq C_T}{\text{arg max }} \left\|\mbf{P}_{{R_s},i} \mbf{x}_R-\mbf{x}_R\right\|_2.
\end{equation}

\noindent
\textbf{Estimation of $\tilde{\mbf{H}}^k_{R,i}$:}
For the simulations, we have adopted the channel estimation technique as described in \cite{A.BabaeiJuly2013}. The estimation is carried out by having all the BSs in the $i^{\text{th}}$ cluster $\mathcal{M}_k$ send training symbols to the radar and the radar utilizes the received signal to estimate $\tilde{\mbf{H}}^k_{R,i}$ and find $\mbf{P}_{R,i}/\mbf{P}_{R_s,i}$. It is assumed that BSs in the $i^{\text{th}}$ cluster  $\mathcal{M}_k$ can cooperate in their choices of training symbols and transmission power. Further assuming channel reciprocity, the channel from $n^\mathrm{th}$ BS to the radar transmitter is $\left({\mbf{H}}^k_{nR,i}\right)^*$. The composite channel from all of the BSs in the cluster to the radar transmitter is given by
\begin{equation}
\bar{\mbf{H}}^k_{R,i}=\left[ \left(\mbf{H}^k_{1R,i}\right)^* \left(\mbf{H}^k_{2R,i}\right)^*\cdots\left(\mbf{H}^k_{M_k R,i}\right)^*\right].
\end{equation}
Obviously, $\tilde{\mbf{H}}^k_{R,i}=\left(\bar{\mbf{H}}^k_{R,i}\right)^*$. The assumption of coordination among BSs reduces $\bar{\mbf{H}}^k_{R,i}$ to a standard MIMO channel. So, by using standard MIMO channel estimation algorithm, the equations for estimation of composite interference channel is as shown below \cite{A.BabaeiJuly2013}
\begin{equation}
\mbf{y}_e=\sqrt{\frac{\rho}{M_k N_{\text{BS}}}} \bar{\mbf{H}}^k_{R,i} \mbf{s}_e +\mbf{w}_e,\quad 1\leq e \leq L_t. 
\end{equation}
Here, $L_t$ is the fixed period at the beginning of each block of $L$ channel uses where the estimation is performed; $\mbf{y}_e$ and $\mbf{w}_e$ are respectively the $M_R$-dimensional received signal vector and  the noise vector at time $e$; and $\rho$ is the average SNR at each receiving antenna. The maximum likelihood (ML) estimation of the $\bar{\mbf{H}}^k_{R,i}$ is given by \cite{A.BabaeiJuly2013}
\begin{equation}
	{\hat{\bar{\mbf{H}}}}^k_{R,i}\left(ML\right)=\sqrt{\frac{M_k N_{\text{BS}}}{\rho}} \mbf{Y} {\mbf{S}}^* \left({\mbf{S}} {\mbf{S}}^*\right)^{-1}
\end{equation}
where the matrices are: ${\mbf{Y}=\begin{bmatrix}\mbf{y}_1 &\mbf{y}_2 &\cdots& \mbf{y}_{L_t}\end{bmatrix}}$, ${\mbf{W}=\begin{bmatrix}\mbf{w}_1&\mbf{w}_2&\cdots& \mbf{w}_{L_t}\end{bmatrix}}$, and ${\mbf{S}=\begin{bmatrix}\mbf{s}_1&\mbf{s}_2&\cdots&\mbf{s}_{L_t}\end{bmatrix}}$. The optimal training symbols that minimizes the mean square error is chosen so that ${\mbf{S}} {\mbf{S}}^*=L_t \mbf{I}_{M_k N_{\text{BS}}}$. Cooperation among the BSs is required to choose the optimal training sequence. So, the estimation of $\tilde{\mbf{H}}^k_{R,i}$ is given by

\begin{equation}
\hat{\tilde{\mbf{H}}}^k_{R,i}=\left[{\hat{\bar{\mbf{H}}}}^k_{R,i}\left(ML\right)\right]^*.	
\end{equation}

\subsection{Precoder Design for Cooperation}
In this mode, the radar communicates with the CoMP system. The radar informs the communication system of the channel matrices corresponding to base station locations that will be affected by its transmission and the locations that will have a suppressed interference due to the null space or small singular value space projection. The cellular network communicates the information about the BS clustering to the radar. The channel estimation is also carried out in this phase. But this cooperation is for brief period to take care of the operational security concerns of the radar. In this phase, radar operates in broadcast (BC) mode on downlink, i.e. from radar to the BSs. Simple linear precoding (LP) techniques such as zero forcing (ZF) and minimum mean-square error (MMSE) linear precoders have been considered for this MIMO BC mode. By inverting the channel matrix at the transmitter, the zero-forcing precoder can completely eliminate the interference among the BSs. However, a price of high transmission energy has 
to be paid especially for near singular matrix channels. This problem can be partially mitigated by using the MMSE linear precoder. This kind of precoder balances the transmission energy and interference level to achieve the minimum detection error.

In this mode, radar signal is given by
\begin{equation}
{\mbf{x}_R = [x_{R,1} ...x_{R,d_R} ]^T} \in \mathbb{C}^{d_R}	
\end{equation}
where, $d_R$ is the number of independent information streams intended for the BSs and $d_R \leq min({M} {N_{\text{BS}}}, M_R )$ and $M$ is the total number of BSs in the communication system.

Using ZF criterion, the precoding matrix is given by \cite{Peel2005}
\begin{equation}
\mbf{P}_R={\tilde{\mbf{H}}}^*_R \left({\tilde{\mbf{H}}}_R {\tilde{\mbf{H}}}^*_R\right)^{-1}
\label{eq_coop} 
\end{equation}
where, ${\tilde{\mbf{H}}}_R$ is the composite channel between the radar and all the BSs in the communication system and is given by
\begin{equation}
\tilde{\mbf{H}}_R={\left[ {\left({\mbf{H}_{1R}}\right)}^*  {\left({\mbf{H}_{2R}}\right)}^*\cdots{\left({\mbf{H}_{MR}}\right)}^* \right]}^*.	
\end{equation}
The inverse in equation (\ref{eq_coop}) can be performed only when the following condition is satisfied
\begin{equation*}
M_R \geq d_R.
\end{equation*}
Similarly, using MMSE criterion, the precoder is as follows \cite{ShaoOctober2007}

\begin{equation}
\mbf{P}_R={\tilde{\mbf{H}}}^*_R \left({\tilde{\mbf{H}}}_R {\tilde{\mbf{H}}}^*_R+ {d_R} {\sigma}^2 \mbf{I}\right)^{-1}. 
\end{equation}

\subsection{The two modes of operation and the PRI of Radar}
The radar will keep switching between these two modes of operation. During the first part of its Pulse Repetition Interval (PRI), it will operate in cooperation mode for a short period. But this short period is very critical for the proper operation of the radar for the rest of the PRI. In this brief period, it will collect all the information about the BS clustering of the CoMP and estimate the composite interference channels between itself and the cellular network. The radar will also broadcast the information it wants to communicate to the CoMP system. Based on the vital information it has gathered in the cooperation mode, it will enter next into the interference-mitigation mode for the rest of the PRI. Then the same cycle would repeat.   
\section{Spectrum Sharing Algorithms}\label{sec:algos}
Based on the background theories presented in the last section about the interference-mitigation mode, we have proposed two new algorithms to implement Switched Small Singular Value Space Projection. As algorithmic implementation of Switched Null Space Projection is already available in literature \cite{A.Khawar, Shajaiah2014, SKA+14DySPAN}, that part is omitted here. 
  
\subsection{Optimal Cluster Selection Algorithm}
Our optimal cluster selection algorithm, shown in Algorithm \eqref{alg:select}, selects the best composite interference channel based on the optimality criteria as mentioned in equations \eqref{eq:crit1} and \eqref{eq:crit2}. Channel State Information (CSI)s are obtained using the channel estimation technique as described in the last section. The SSVSP projection matrices of all the clusters, available at that time interval are then found using Algorithm (\ref{alg:ssvsp}). Algorithm (\ref{alg:ssvsp}) also calculates the difference between the precoded and the original radar signal and returns it to Algorithm (\ref{alg:select}). Once Algorithm (\ref{alg:select}) receives the difference metrics of all the composite interference channels, it selects the best composite interference channel associated with the optimal cluster and sends it to Algorithm (\ref{alg:ssvsp}) for SSVSP radar signal. 
 
\begin{algorithm}
\caption{Optimal Cluster Selection Algorithm}\label{alg:select}
\begin{algorithmic}
\LOOP
	\FOR{$i=1:C_T$}
		\STATE{Estimate CSI of $\tilde{\mbf{H}}^k_{R,i}$.}
		\STATE{Send $\tilde{\mbf{H}}^k_{R,i}$ to Algorithm \eqref{alg:ssvsp} for small singular value space computation.}
		\STATE{Receive $\left\|\mbf{P}_{{R_s,i}} \mbf{x}_R-\mbf{x}_R\right\|_2$ from  Algorithm \eqref{alg:ssvsp}.}
	\ENDFOR
	\STATE{Find {$i_{min}=\underset{1 \leq i\leq C_T}{\text{arg min }} \left\|\mbf{P}_{{R_s,i}} \mbf{x}_R-\mbf{x}_R\right\|_2$.}}
\STATE{Set $\breve{\mbf H} = (\tilde{\mbf{H}}^k_{R,i})_{i_{\text{min}}}$ as the best composite interference channel associated with the optimal cluster.}
\STATE{Send $\breve{\mbf H}$ to Algorithm \eqref{alg:ssvsp} to get SSVSP radar waveform.}
\ENDLOOP
\end{algorithmic}
\end{algorithm}

\subsection{Small Singular Value Space Projection (SSVSP) Algorithm}
In this section, we present our proposed Small Singular Value Space Projection algorithm. On the first cycle, Algorithm (\ref{alg:ssvsp}) gets the CSI estimates of the composite interference channels from Algorithm (\ref{alg:select}) and finds the corresponding small singular value space projection matrices using the singular value decomposition (SVD) theorem. It also calculates the associated difference metrics and return them to Algorithm (\ref{alg:select}). On the second cycle, after receiving the best composite interference channel matrix associated with the optimal cluster from Algorithm (\ref{alg:select}), it finally calculates the precoded SSVSP radar signal by performing another SVD.

\begin{algorithm}
\caption{Small Singular Value Space Projection (SSVSP) Algorithm}\label{alg:ssvsp}
\begin{algorithmic}
\IF {$\tilde{\mbf{H}}^k_{R,i}$ received from Algorithm \eqref{alg:select}} 
	\STATE{Perform SVD on $\tilde{\mbf{H}}^k_{R,i}$ (i.e. $\tilde{\mbf{H}}^k_{R,i}=\tilde{\mbf{U}}_i \tilde{\mbf{S}}_i {\tilde{\mbf{V}}^*_i}$).}
		\STATE{Find small singular value space projection matrix $\mbf{P}_{{R_s,i}}=\bar{\mbf{V}}_{s,i} {\bar{\mbf{V}}}^*_{s,i}$.}
		\STATE{Calculate $\left\|\mbf{P}_{{R_s,i}} \mbf{x}_R-\mbf{x}_R\right\|_2$.}					
		\STATE{Send $\left\|\mbf{P}_{{R_s,i}} \mbf{x}_R-\mbf{x}_R\right\|_2$ to Algorithm \eqref{alg:select}.}		
				
		\ENDIF

\IF {$\breve{\mbf H}$ received from Algorithm \eqref{alg:select}}
		\STATE{Perform SVD on $\breve{\mbf{H}}$ (i.e. $\breve{\mbf{H}}=\breve{\mbf{U}} \breve{\mbf{S}} {\breve{\mbf{V}}^*}$).}
		\STATE{Find small singular value space projection matrix $\breve{\mbf{P}}_{R_s}=\breve{\bar{\mbf{V}}}_s \breve{{\bar{\mbf{V}}}}^*_s$.}
	\STATE{Calculate SSVSP radar signal $\breve{\mbf{x}}_{R_s}=\breve{\mbf{P}}_{R_s} \mbf{x}_{R}$.}			
\ENDIF 
\end{algorithmic}
\end{algorithm}

\section{Simulation Results}\label{sec:sim}

In this section, the performance of MIMO radar is compared in terms of CRB for target direction estimation with and without radar precoder and as a function of number of antennas per BS, number of BSs per cluster, and number of radar antennas with either SNSP or SSVSP. We have also explored the effect of null space estimation error on target direction estimation and radar interference to clusters. The distance of target to radar array and the radar inter-element spacing are assumed to be $r_0 = 5$ km and $3\lambda/4$, respectively. The frequency of operation is 3.5 GHz. The signal-to-noise ratio between the radar and its target is denoted by SNR while the same parameter between the radar and the cluster of BSs of CoMP system is represented by $\rho$. The target direction is assumed to be at $\theta = 0^\circ$.

\subsection{Performance Analysis of Interference Mitigating Precoder}

\begin{figure}[!t]
\centering
\includegraphics[width=\linewidth]{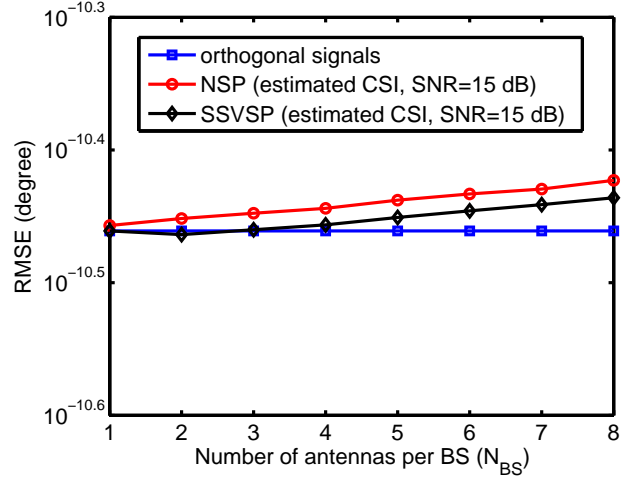}
\caption{CRB on direction estimation of the target as a function of antennas employed by the BS when using NSP and SSVSP in a spectrum sharing scenario between MIMO radar and cellular system. A cluster size of $M_k=3$ is used while having $M_R=100$ radar antenna elements.}
\label{nsp_vs_ssvsp_100}
\end{figure}

Fig. (\ref{nsp_vs_ssvsp_100}) shows the radar CRB performance for orthogonal radar signals as well as precoded radar signals for NSP and SSVSP with estimated Channel State Information (CSI). The figure reflects the fact that SSVSP performs better than NSP from the perspective of radar target detection capability as predicted. This plot also shows that as $N_{\text{BS}}$ increases, the target localization performance of radar degrades. Because with the increase in $N_{\text{BS}}$, $\left(M_R-M_k N_{\text{BS}}\right)$ decreases for constant $M_R \text{ and } M_k$. So, null space of $\tilde{\mbf{H}}^k_{R,i}$ shrinks with consequent impact on the precoder which in turn degrades CRB$(\theta)$.

\begin{figure}[!t]
\centering
\includegraphics[width=\linewidth]{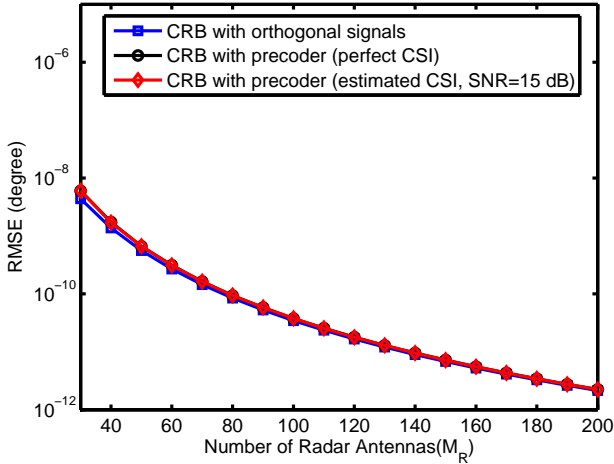}
\caption{Effect of number of radar antennas $(M_R)$ on the CRB($\theta$) performance of radar. A cluster size of $M_k=3$ is used and $N_{BS}=8$.}
\label{fig9}
\end{figure}

In Fig. (\ref{fig9}), we have investigated the effect of number of radar antennas $(M_R)$ on the CRB performance of radar. As obvious from equation (\ref{eq:crb}) that CRB explicitly depends on $M_R$ and improves when $M_R$ increases. For precoded radar signals, increase in $M_R$ leads to increase in the nullity of $\tilde{\mbf{H}}^k_{R,i}$ which impacts the choice of $\mbf{P}_{R,i}$ and results in improved CRB performance. Results in this plot indicate that increasing the number of radar antennas can compensate for the performance degradation in target direction estimation due to correlation in the precoded radar signals. Consequently, for a given target RMSE, number of radar antennas must increase when radar employs a precoder to zero-force or minimize its interference at the clusters. 

\begin{figure}[!t]
\centering
\includegraphics[width=\linewidth]{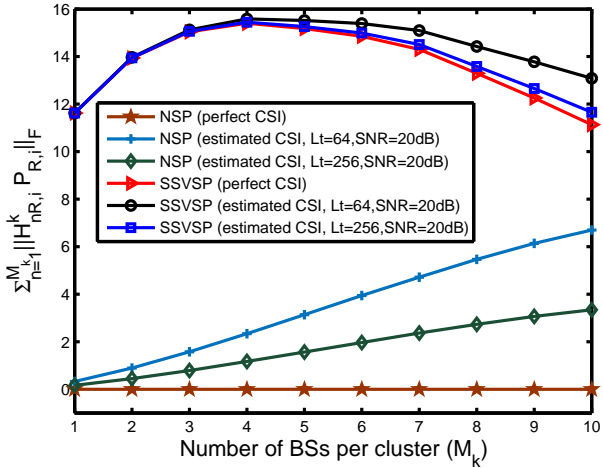}
\caption{Variation of radar interference on BS clusters with cluster size ($M_k$). $N_{\text{BS}}=6$ BS antenna elements are used while having $M_R=100$ radar antenna elements.}
\label{fig8}
\end{figure}

\begin{figure}[!t]
\centering
\includegraphics[width=\linewidth]{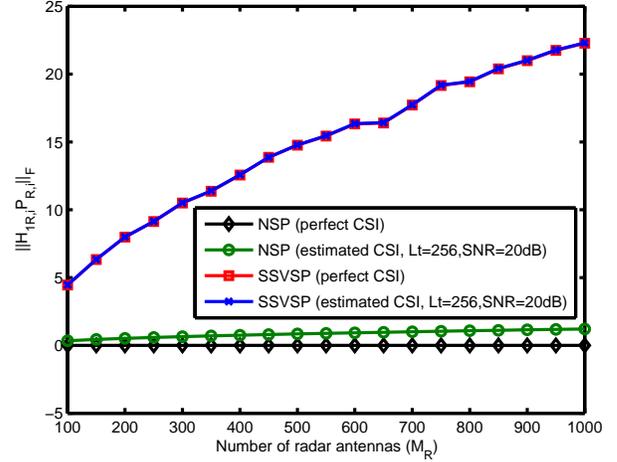}
\caption{ Radar Interference on one BS with number of radar antenna elements, $M_R$. $N_{\text{BS}}=8$ antenna elements per BS are used while having cluster size, $M_k=3$.}
\label{Fig7}
\end{figure}

\begin{figure}[!t]
\centering
\includegraphics[width=\linewidth]{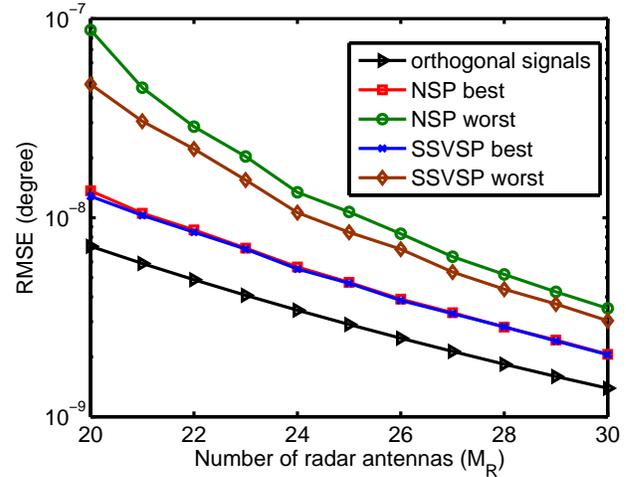}
\caption{CRB on direction estimation of the target as a function of number of antenna elements employed by the radar ($M_R$) with switched NSP and SSVSP while in a spectrum sharing scenario between MIMO radar and cellular system. A cluster size of $M_k=3$ is used while having estimated CSI. Radar beam is swept across 4 clusters ($C_T=4$) and the nullity of the clusters is varied by varying the number of antenna elements per BS, [$N_{\text{BS}}=(6,5,4,3)$].}
\label{Fig8}
\end{figure}

In Fig. (\ref{fig8}), the effect of interference at the clusters is measured using the metric, ${\sum^{M_k}_{n=1}\left\|{\mbft{H}_{nR,i}^k} {\mbf{P}_{R,i}}\right\|}_F$. Interference becomes less significant as the duration of training phase, $L_t$ increases while it will be completely eliminated with perfect knowledge of $\tilde{\mbf{H}}^k_{R,i}$. So, estimated CSI approaches perfect CSI as $L_t$ increases. This plot also reflects the fact that from the perspective of interference at the clusters, NSP outperforms SSVSP to a great extent. Fig. (\ref{Fig7}) represents the same facts for a specific BS in a cluster. Fig. (\ref{Fig8}) plots radar target estimation capability with switched NSP and SSVSP. It is obvious that both of the beam sweeping methods improve radar performance when the beam is projected on to the optimal cluster associated either with the maximum null space or minimum deviation of the precoded radar signal from the original one. For the worst-case scenarios, SSVSP outperforms SNSP.
    
\subsection{Performance Analysis of Information Exchange Precoder}

\begin{figure}[!t]
\centering
\includegraphics[width=\linewidth]{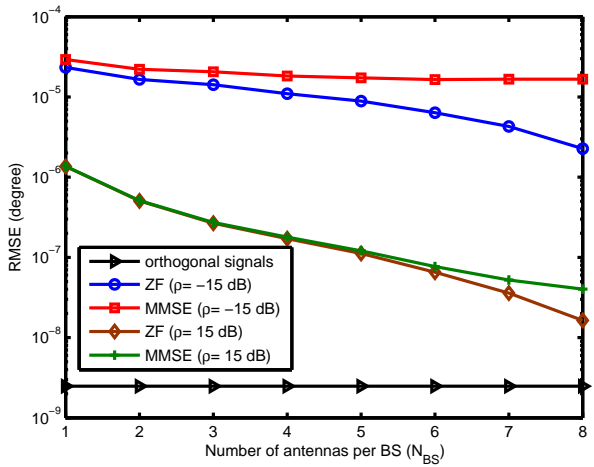}
\caption{CRB on direction estimation of the target as a function of the number of antenna elements employed by the BS while in a cooperative spectrum sharing scenario between MIMO radar and cellular system. A total of 3 BSs ($M=3$) is used while having estimated CSI and $M_R=26$ radar antenna elements.}
\label{Fig9}
\end{figure}

\begin{figure}[!t]
\centering
\includegraphics[width=\linewidth]{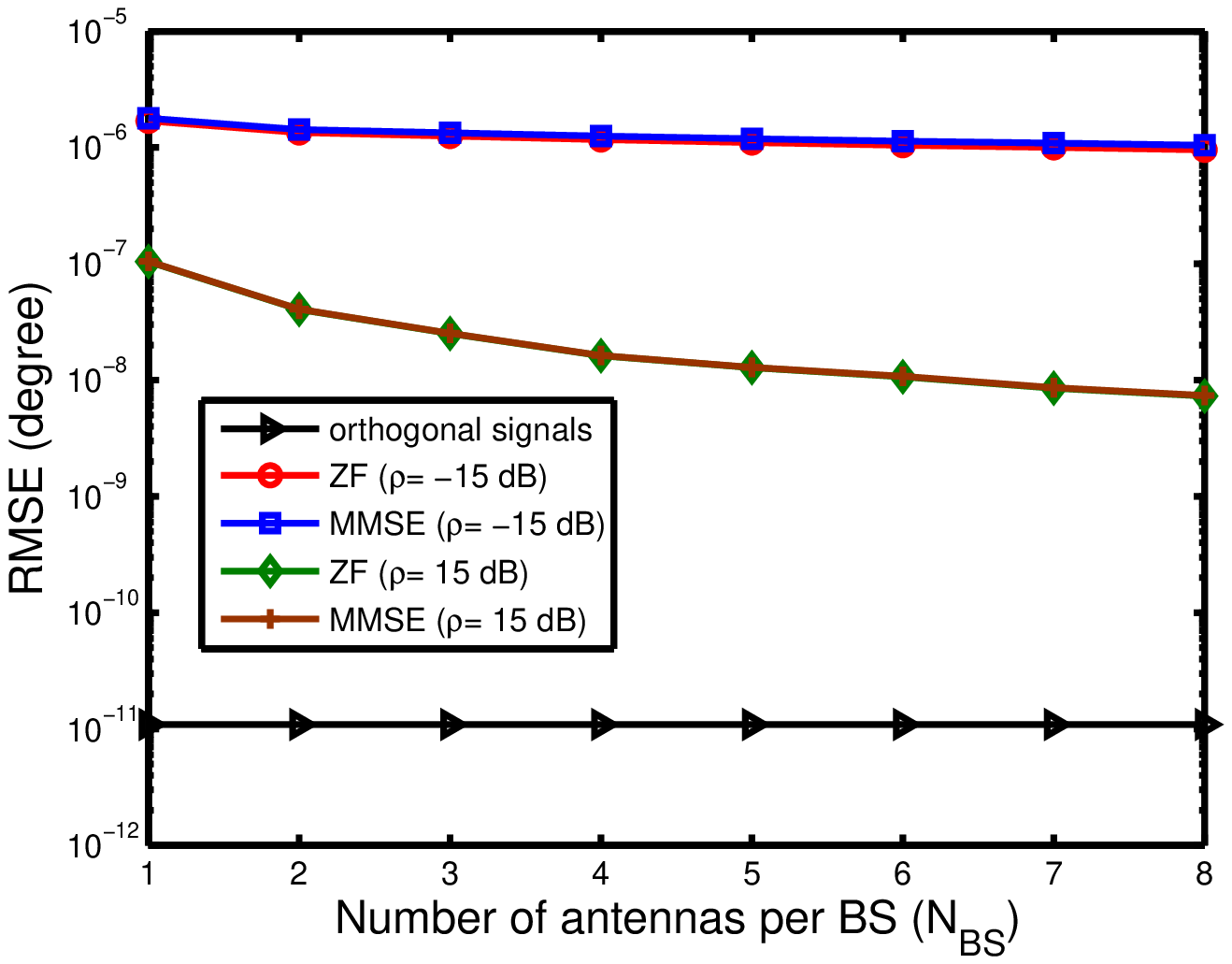}
\caption{CRB on direction estimation of the target as a function of the number of antenna elements employed by the BS while in a cooperative spectrum sharing scenario between MIMO radar and cellular system. A total of 3 BSs ($M=3$) is used while having estimated CSI and $M_R=100$ radar antenna elements.}
\label{coop_crb_100}
\end{figure}

\begin{figure}[!t]
\centering
\includegraphics[width=\linewidth]{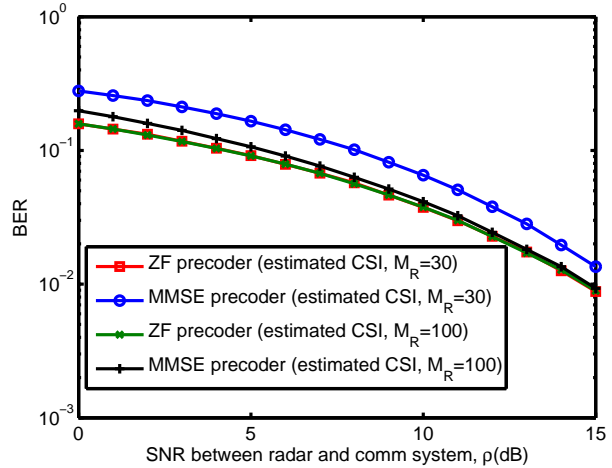}
\caption{BER performance during cooperation mode using QPSK waveform. A total of 4 BSs ($M=4$) is used.}
\label{coop_ber}
\end{figure}

Fig. (\ref{Fig9}) through Fig. (\ref{coop_ber}) show the spectrum sharing scenario during cooperation mode. Fig. (\ref{Fig9}) and Fig. (\ref{coop_crb_100}) show that target location performance of both ZF and MMSE precoding is same at high $\rho$ for low values of $N_{\text{BS}}$ while they start deviating from each other at high $N_{\text{BS}}$. ZF precoding outperforms MMSE precoding at low $\rho$. They also reflect the fact that this discrepancy in performance can be overcome by increasing the number of radar antenna elements. Fig. (\ref{coop_ber}) represents the performance of the cluster in terms of BER. As predicted before, ZF outperforms MMSE precoding in this case too, but at the expense of high transmission energy. The increase in the number of radar antenna is the obvious solution to close the gap between the performance of these two precoding methods.

\section{Conclusion}\label{sec:conc}
In this paper, the authors have designed precoders of a radar in a scenario where a MIMO radar co-exists spectrally with a CoMP commercial communication system. We have investigated the design of precoder for two modes of radar operation: interference-mitigation mode, when the radar attempts to avoid interference with the communication system and cooperation mode, when it exchanges information with the CoMP system. For the interference-mitigation mode, the radar steers its beam towards the optimal clusters of base stations and keeps switching the beam from one optimal cluster to another. In case of traditional SNSP, the optimality is decided upon the maximum nullity of the cluster. A new space projection and switching method has been proposed where the beam is steered to the small singular value space of the optimal cluster whose optimality is chosen based on the minimum difference between the precoded and original radar signal. For cooperation mode, the precoder is designed based on the conventional zero-
forcing and MMSE criteria to reduce BER at the BSs for effective detection. Channel estimation error has been taken into account for both modes of operation. Although the introduction of precoding tries to mitigate the radar interference at the clusters during interference-free mode and minimize interference among BSs during cooperation mode, in doing so it causes the radar probing signals loose their orthogonality with consequent performance loss in radar's target parameter estimation capability. SSSVSP mitigates that effect to some extent but increase in the number of radar antenna elements has stronger impact.


\bibliographystyle{ieeetr}
\bibliography{IEEEabrv,clustering}

\begin{thebibliography}{10}

\bibitem{url_1}
{http://www.qualcomm.com/solutions/wireless-networks/technologies/1000x-data.}

\bibitem{Andrews2007}
J.~G. Andrews, W.~Choi, and R.~W.~H. Jr, ``Overcoming interference in spatial
  multiplexing {MIMO} cellular networks,'' {\em IEEE Wireless Commun. Mag.},
  vol.~14, pp.~95--104, December 2007.

\bibitem{LTE_Rel11}
{4G Americas}, ``{4G} mobile broadband evolution: {3GPP Release 11 \& Release
  12} and beyond,'' 2014.

\bibitem{Memo2010}
{Presidential Memorandum}, ``Unleashing the wireless broadband revolution,''
  2010.

\bibitem{NTIA10}
{National Telecommunications and Information Administration (NTIA)}, ``An
  assessment of the near-term viability of accommodating wireless broadband
  systems in the 1675-1710 {MH}z, 1755-1780 {MH}z, 3500-3650 {MH}z, 4200-4220
  {MH}z, and 4380-4400 {MH}z bands ({F}ast {T}rack {R}eport).'' Online, October
  2010.

\bibitem{NTIA12}
{National Telecommunications and Information Administration (NTIA)}, ``Analysis
  and resolution of {RF} interference to radars operating in the band 2700-2900
  {MH}z from broadband communication transmitters.'' Online, October 2012.

\bibitem{FCC12_SmallCells}
{Federal Communications Commission (FCC)}, ``{FCC} proposes innovative small
  cell use in {3.5 GHz} band.'' Online:
  http://www.fcc.gov/document/fcc-proposes-innovative-small-cell-use-35-ghz-ba%
nd, December 12, 2012.

\bibitem{FCC_5GHz_Radar06}
{Federal Communications Commission (FCC)}, ``In the matter of revision of parts
  2 and 15 of the commission�s rules to permit unlicensed national
  information infrastructure {(U-NII)} devices in the 5 {GH}z band.'' MO\&O, ET
  Docket No. 03-122, June 2006.

\bibitem{S.SodagariDec.2012}
S.~Sodagari, A.~Khawar, T.~C. Clancy, and R.~McGwier, ``{A} projection based
  approach for radar and telecommunication systems coexistence,'' {\em Global
  Communications Conference (GLOBECOM), 2012 IEEE}, pp.~5010--5014, Dec. 2012.

\bibitem{A.Khawar}
A.~Khawar, A.~Abdel-Hadi, and T.~C. Clancy, ``Spectrum sharing between {S-band}
  radar and {LTE} cellular system: {A} spatial approach,'' in {\em 2014 IEEE
  International Symposium on Dynamic Spectrum Access Networks: SSPARC Workshop
  (IEEE DySPAN 2014 - SSPARC Workshop)}, (McLean, USA), Apr. 2014.

\bibitem{Babaei2013}
A.~Babaei, W.~H. Tranter, and T.~Bose, ``{A} nullspace-based precoder with
  subspace expansion for radar/communications coexistence,'' {\em Globecom 2013
  - Signal Processing for Communications Symposium}, 2013.

\bibitem{WMW+08}
L.~S. Wang, J.~P. McGeehan, C.~Williams, and A.~Doufexi, ``Application of
  cooperative sensing in radar-communications coexistence,'' {\em IET
  Communications}, vol.~2, pp.~856--868, 2008.

\bibitem{BNR12}
S.~S. Bhat, R.~M. Narayanan, and M.~Rangaswamy, ``Bandwidth sharing and
  scheduling for multimodal radar with communications and tracking,'' in {\em
  IEEE Sensor Array and Multichannel Signal Processing Workshop}, p.~233–236,
  2012.

\bibitem{SPC12}
R.~Saruthirathanaworakun, J.~Peha, and L.~Correia, ``Performance of data
  services in cellular networks sharing spectrum with a single rotating
  radar,'' in {\em IEEE International Symposium on a World of Wireless, Mobile
  and Multimedia Networks (WoWMoM)}, pp.~1--6, 2012.

\bibitem{GhorbanzadehMilcom2014}
M.~Ghorbanzadeh, A.~Abdelhadi, and T.~C. Clancy, ``A utility proportional
  fairness resource allocation in spectrally radar-coexistent cellular
  networks,'' in {\em IEEE Military Communications Conference (MILCOM)},
  October 2014.

\bibitem{SKA+14DySPAN}
H.~Shajaiah, A.~Khawar, A.~Abdel-Hadi, and T.~C. Clancy, ``Resource allocation
  with carrier aggregation in {LTE Advanced} cellular system sharing spectrum
  with {S-band} radar,'' in {\em IEEE International Symposium on Dynamic
  Spectrum Access Networks: SSPARC Workshop (IEEE DySPAN 2014 - SSPARC
  Workshop)}, (McLean, USA), Apr. 2014.

\bibitem{FHH14}
M.~P. Fitz, T.~R. Halford, I.~Hossain, and S.~W. Enserink, ``Towards
  simultaneous radar and spectral sensing,'' in {\em IEEE International
  Symposium on Dynamic Spectrum Access Networks (DYSPAN)}, pp.~15--19, April
  2014.

\bibitem{PMM+14}
F.~Paisana, J.~P. Miranda, N.~Marchetti, and L.~A. DaSilva, ``Database-aided
  sensing for radar bands,'' in {\em IEEE International Symposium on Dynamic
  Spectrum Access Networks (DYSPAN)}, pp.~1--6, April 2014.

\bibitem{Shahriar_WCNC}
C.~Shahriar, A.~Abdelhadi, and T.~Clancy, ``{Overlapped-MIMO Radar Waveform
  Design for Coexistence With Communication Systems},'' {\em Under Submission}.

\bibitem{DH13}
H.~Deng and B.~Himed, ``Interference mitigation processing for spectrum-sharing
  between radar and wireless communications systems,'' {\em IEEE Transactions
  on Aerospace and Electronic Systems}, vol.~49, no.~3, pp.~1911--1919, 2013.

\bibitem{A.BabaeiJuly2013}
A.~Babaei, W.~H. Tranter, and T.~Bose, ``{A} practical precoding approach for
  radar/communications spectrum sharing,'' {\em Cognitive Radio Oriented
  Wireless Networks (CROWNCOM)}, pp.~13--18, July 2013.

\bibitem{KAC14_Milcom}
A.~Khawar, A.~Abdel-Hadi, and T.~C. Clancy, ``On the impact of time-varying
  interference-channel on the spatial approach of spectrum sharing between
  {S-band} radar and communication system,'' in {\em IEEE Military
  Communications Conference (MILCOM)}, 2014.

\bibitem{Negro2010}
F.~Negro, S.~P. Shenoy, I.~Ghauri, and D.~Slock, ``{O}n the {MIMO} interference
  channel,'' {\em Proc. of ITA}, pp.~1--9, 2010.

\bibitem{S.KavianiDec.2011}
S.~Kaviani, O.~Simeone, W.~A. Krzymien, and S.~Shamai, ``{L}inear {MMSE}
  precoding and equalization for network {MIMO} with partial cooperation,''
  {\em IEEE Global Communications Conference (GLOBECOM)}, pp.~1--6, Dec. 2011.

\bibitem{Boccardi2007}
F.~Boccardi and H.~Huang, ``{L}imited downlink network coordination in cellular
  networks,'' {\em IEEE 18th International Symposium on Personal, Indoor and
  Mobile Radio Communications (PIMRC 2007)}, pp.~1--5, September 2007.

\bibitem{MF11}
P.~Marsch and G.~P. Fettweis, eds., {\em {Coordinated Multi-Point in Mobile
  Communications: From Theory to Practice}}.
\newblock Cambridge University Press, 2011.

\bibitem{Papadogiannis2008}
A.~Papadogiannis, D.~Gesbert, and E.~Hardouin, ``{A} dynamic clustering
  approach in wireless networks with multi-cell cooperative processing,'' {\em
  IEEE International Conference on Communications}, pp.~4033 -- 4037, 2008.

\bibitem{Boccardi2008}
F.~Boccardi, H.~Huang, and A.~Alexiou, ``{N}etwork {MIMO} with reduced backhaul
  requirements by {MAC} coordination,'' {\em IEEE 42nd Asilomar Conference on
  Signals, Systems and Computers (ASILOMAR 2008)}, pp.~1125--1129, October
  2008.

\bibitem{Papadogiannis2010}
A.~Papadogiannis and G.~C. Alexandropoulos, ``{T}he value of dynamic clustering
  of base stations for future wireless networks,'' {\em IEEE International
  Conference on Fuzzy Systems}, pp.~1--6, July 2010.

\bibitem{KAC+14ICNC}
A.~Khawar, A.~Abdel-Hadi, T.~C. Clancy, and R.~McGwier, ``Beampattern analysis
  for {MIMO} radar and telecommunication system coexistence,'' in {\em IEEE
  International Conference on Computing, Networking and Communications, Signal
  Processing for Communications Symposium (ICNC'14 - SPC)}, 2014.

\bibitem{KAC14_TDetect}
A.~Khawar, A.~Abdelhadi, and T.~C. Clancy, ``Target detection performance of
  spectrum sharing {MIMO} radars,'' {\em arXiv:1408.0540}.

\bibitem{LS08}
J.~Li and P.~Stoica, {\em {MIMO} Radar Signal Processing}.
\newblock Wiley-IEEE Press, 2008.

\bibitem{KAC14DySPANWaveform}
A.~Khawar, A.~Abdel-Hadi, and T.~C. Clancy, ``{MIMO} radar waveform design for
  coexistence with cellular systems,'' in {\em 2014 IEEE International
  Symposium on Dynamic Spectrum Access Networks: SSPARC Workshop (IEEE DySPAN
  2014 - SSPARC Workshop)}, (McLean, USA), Apr. 2014.

\bibitem{KAC14_QPSK}
A.~Khawar, A.~Abdelhadi, and T.~C. Clancy, ``{QPSK} waveform for {MIMO} radar
  with spectrum sharing constraints,'' {\em arXiv:1407.8510}.

\bibitem{CadambeAug2008}
V.~R. Cadambe and S.~A. Jafar, ``{I}nterference {A}lignment and {D}egrees of
  {F}reedom of the {K}-{U}ser {I}nterference {C}hannel,,'' {\em IEEE Trans.
  Inf. Theory}, vol.~54, no. 8, pp.~3425--3441, Aug, 2008.

\bibitem{Kalm1996}
D.~Kalman, ``{A} {S}ingularly {V}aluable {D}ecomposition: {T}he {SVD} of a
  {M}atrix,'' {\em The College Mathematics Journal}, vol.~27, January 1996.

\bibitem{Jafar2011}
S.~A. Jafar, ``{I}nterference alignment: {A} new look at signal dimensions in a
  communication network,'' {\em {F}oundations and {T}rends in {C}ommunications
  and {I}nformation {T}heory}, vol.~7, no. 1, 2011.

\bibitem{Peel2005}
C.~Peel, B.~Hochwald, and A.~Swindlehurst, ``{A} {v}ector-{p}erturbation
  {t}echnique for {n}ear-{c}apacity {m}ultiantenna {m}ultiuser
  {c}ommunication-part i: {c}hannel {i}nversion and {r}egularization,'' {\em
  IEEE Trans. Comm.}, vol.~53, p.~195 – 202, 2005.

\bibitem{ShaoOctober2007}
X.~Shao, J.~Yuan, and Y.~Shao, ``{E}rror performance analysis of linear zero
  forcing and {MMSE} precoders for {MIMO} broadcast channels,'' {\em IET
  Commun.}, vol.~1, no. 5, pp.~1067--1074, October, 2007.

\bibitem{Shajaiah2014}
H.~Shajaiah, A.~Abdelhadi, and C.~Clancy, ``{I}mpact of radar and communication
  coexistence on radar's detectable target parameters,'' {\em Under
  Submission}, 2014.

\end{thebibliography}
\end{document}